\documentclass[a4paper,11pt]{article}

\usepackage{amsmath,amssymb}
\usepackage{color}
\usepackage{cite}

\newtheorem{theorem}{Theorem}[section]

\def\ri{\mathrm{i}}\def\rd{\mathrm{d}}
\def\tr{\mbox{ \rm tr\,}} \newcommand{\ima}{\mathrm{Im}}
\def\ad{\mbox{\rm ad\,}} 
\def\mod{\mbox{\rm mod\,}}\def\Ad{\mbox{\rm Ad\,}}
\def\diag{\mbox{\rm diag\,}}\def\id{\mbox{\rm id\,}}

\def\g{\mathfrak{g}\,}
\def\SU{\mbox{\rm SU\,}}

\def\asl{\mbox{\rm sl\,}}

\def\su{\mbox{\rm su\,}}

\def\id{\mbox{\rm id\,}}

\def\Res{\mathop{\mbox{\rm Res}\,}\limits}
\def\pd#1,#2{\frac{\partial#1}{\partial#2}}
\def\v#1{\overline{#1}}
\def\d#1,#2{\frac{d#1}{d#2}}
\def\qed{\hbox{${\vcenter{\vbox{\hrule height 0.4pt\hbox{\vrule width 0.4pt height 6pt
\kern 5pt\vrule width 0.4pt}\hrule height 0.4pt}}}$}}

\makeindex 
            
\begin{document}

\title{Hermitian and Pseudo-Hermitian Reduction\\
of the GMV Auxiliary System. Spectral Properties\\ of the Recursion Operators}
\author{A. B. Yanovski$^1$ and T. I. Valchev$^2$\\
\small $^1$ Department of Mathematics \& Applied Mathematics,\\
\small University of Cape Town, Rondebosch 7700,
Cape Town, South Africa\\
\small $^2$ Institute of Mathematics and Informatics,\\
\small Bulgarian Academy of Sciences, Acad. G. Bonchev Str., 1113 Sofia, Bulgaria\\
\small E-mails: Alexandar.Ianovsky@uct.ac.za,\quad tiv@math.bas.bg}
\date{}
\maketitle

\abstract{We consider simultaneously two different reductions of a Zakharov-Shabat's spectral problem in pole gauge. Using the concept of gauge equivalence, we construct expansions over the eigenfunctions of the recursion operators related to the afore-mentioned spectral problem with arbitrary constant asymptotic values of the potential functions. In doing this, we take into account the discrete spectrum of the scattering operator. Having in mind the applications to the theory of the soliton equations associated to the GMV systems, we show how these expansions modify depending on the symmetries of the functions we expand.}

\section{Introduction. The GMV System}
\label{sec:1}

We are going to study the auxiliary linear problem
\begin{eqnarray}\label{eq:gsys}
&&\tilde{L}^0\psi=(\ri\partial_x - \lambda S)\psi=0,\qquad \lambda\in\mathbb{C}, \qquad S = \left(\begin{array}{ccc} 0&u&v\\ \epsilon u^*&0&0\\v^*&0&0
\end{array}\right), \quad \epsilon=\pm 1
\end{eqnarray}
and the theory of expansions over its adjoint solutions. In the above, the potential functions $(u,v)$ are smooth complex valued functions on $x\in\mathbb{R}$ and $^*$ stands for the complex conjugation. In addition, $u$ and $v$ satisfy the relations:
\begin{equation}\label{eq:uvcon}
\epsilon |u|^2+|v|^2=1, \qquad \lim_{x\to\pm\infty} u(x)=u_{\pm}, \qquad \lim_{x\to\pm\infty} v(x)=v_{\pm}\, .
\end{equation}
We shall call \eqref{eq:gsys} $\rm{GMV}_{\epsilon}$ system or $\rm{GMV}_{\pm}$ system\footnote{A more general system was derived independently by Golubchik and Sokolov \cite{golsok}.}. Thus, $\rm{GMV}_{+}$ is the original Gerdjikov-Mikhailov-Valchev system \cite{GMV1} obtained after putting $\epsilon=+1$ in \eqref{eq:gsys}. 

As demonstrated in  \cite{GMV1,GMVSIGMA,YanVal2017}, the $\rm GMV_{\pm}$ system arises naturally when one looks for integrable systems whose Lax operators are subject to  Mikhailov-type reductions. Indeed, the Mikhailov reduction group $G_0$ \cite{Mikh1, Mikh2} acting on the fundamental solutions of \eqref{eq:gsys} is generated by $g_1$ and $g_2$ defined in the following way:
\begin{equation}\label{eq:MG}
\begin{array}{l}
g_1(\psi)(x,\lambda)=\left[Q_{\epsilon}\psi(x,\lambda^{*})^{\dag}Q_{\epsilon}\right]^{-1},\qquad Q_{\epsilon}=\diag(1,\epsilon,1)\; ,\\[4pt]
g_2(\psi)(x,\lambda)=H\psi(x,-\lambda)H \; , \qquad H=\diag(-1,1,1)
\end{array}
\end{equation}
where $\psi$ is any fundamental solution to \eqref{eq:gsys} and $\dag$ denotes Hermitian conjugation. Since $g_1g_2=g_2g_1$ and $g_1^2=g_2^2=\id$, we have that $G_0=\mathbb{Z}_2\times \mathbb{Z}_2$. Reduction conditions \eqref{eq:MG} will be called Hermitian when $\epsilon =1$ and pseudo-Hermitian when $\epsilon =-1$. The requirement that $G_0$ is a reduction group for the soliton equations related to \eqref{eq:gsys} implies that the coefficients of $\tilde{L}^0$ and the coefficients of the $A$-operators 
\begin{equation}\label{eq:LApairs}
\tilde{A}=\ri\partial_t + \sum_{k=0}^n\lambda^k \tilde{A}_k \; ,\qquad \tilde{A}_k\in \asl(3,\mathbb{C})\; ,
\end{equation}
forming $L-A$ pairs for these soliton equations, must satisfy:
\begin{equation}\label{eq:Arestr}
\begin{array}{l}
HSH = -S,\qquad H\tilde{A}_kH = (-1)^{k}\tilde{A}_k\, ,\\[4pt]
Q_{\epsilon}S^{\dag}Q_{\epsilon} = S,\qquad Q_{\epsilon}\tilde{A}_{k}^{\dag}Q_{\epsilon} = \tilde{A}_{k}\, .
\end{array}
\end{equation}

It can be checked that  $S$ is diagonalizable, indeed, one has
\begin{equation}
g^{-1}Sg=J_0
\label{orbit}\end{equation}
where
\begin{eqnarray}\label{eq:g-tr}
g=\frac{1}{\sqrt{2}}\left(\begin{array}{ccc} 1&0&-1\\ \epsilon u^*&\sqrt{2}v& \epsilon u^*\\v^*&-\sqrt{2}u&v^*\end{array}\right), \qquad J_0=\diag(1,0-1)\, .
\end{eqnarray}
Following \cite{YanVal2017}, we shall write $SU(\epsilon)$ referring to $SU(3)$ when $\epsilon = 1$ and $SU(2,1)$ when $\epsilon = -1$. Similar convention will apply to the corresponding Lie algebras.

Since $g(x)\in SU(\epsilon)$ (\ref{orbit}) means that $S(x)$ belongs to the adjoint representation orbit
\[\mathcal{O}_{J_0}(\SU(\epsilon)) := \{ \tilde{X}\in\ri\su(\epsilon) : \tilde{X} = gJ_0g^{-1},\quad g\in\SU(\epsilon)\}\]
of $\SU(\epsilon)$ passing through $J_0$. 

Our approach to the $\rm GMV_{\pm}$ system will be based on its gauge equivalence to a generalized Zakharov-Shabat auxiliary system  (GZS system) on the algebra $\asl(3,\mathbb{C})$. The auxiliary system 
\begin{equation}\label{eq:CBC}
L\psi=\left( \ri\partial_x + q(x) -\lambda J \right) \psi = 0, \qquad \lambda\in\mathbb{C}
\end{equation}
where $q(x)$ and $J$ belong to some irreducible representation of a simple Lie algebra $\g$ is called generalized Zakharov-Shabat system (for that representation of $\g$) in canonical gauge. The element $J$ must be such that the kernel of $\ad_J(.):= [J, .]$ is a Cartan subalgebra
$\mathfrak{h}_J\subset\g$ while $q(x)$ belongs to the orthogonal complement $\mathfrak{h}_J^{\perp}$ of $\mathfrak{h}_J$ with respect to the Killing form:
\begin{equation}\label{eq:Killing}
\langle X, Y \rangle = \tr(\ad_X \ad_Y),\qquad X, Y \in \g\, .
\end{equation}
It is also assumed that the smooth function $q(x)$ vanishes sufficiently fast as $x\to\pm\infty$. System \eqref{eq:CBC} is gauge equivalent to the system 
\begin{equation}\label{eq:BCGgauge}
\tilde{L}\tilde{\Psi} = \left(\ri\partial_x - \lambda
S(x) \right) \tilde{\Psi} = 0, \qquad S(x) \in {\cal O}_J(G)
\end{equation}
where $G$ is the Lie group corresponding to $\g$. Usually it is also required that
\[\lim_{x\to \pm\infty} S(x) = J\]
where the convergence is sufficiently fast but as we shall see in our case it will be different. The concept of gauge transformation and gauge equivalent auxiliary problems was applied for the first time in the case of the Heisenberg ferromagnet equation \cite{ZaTakh79} and its gauge equivalent --- the nonlinear Schr\"odinger equation.
Later, the integrable hierarchies, the conservation laws and Hamiltonian structures associated with \eqref{eq:CBC} and \eqref{eq:BCGgauge} have extensively been studied by using the so-called gauge-covariant theory of the recursion operators related to the GZS systems in canonical and pole gauge \cite{GerYan85,G86,Yan87}. That approach provides a generalization of classical AKNS approach \cite{AKNS}. We recommend the monograph book \cite{GerViYa2008} for further reading.

So for GZS system in pole gauge most of the essential issues could be reformulated from the canonical gauge. The main difficulty is technical --- to explicitly express all the quantities depending on $q$ and its derivatives through $S$ and its derivatives. A clear procedure of how to do that is described in \cite{Yan87}.  The reader who is interested in that subject can find more details in \cite{Yan93} regarding GZS related to $\asl(3,\mathbb{C})$ with no reductions imposed and in \cite{YanVi2012SIGMA} regarding the geometry of the recursion operators for $\asl(3,\mathbb{C})$ in general position. We also refer to \cite{YanVi2012JNMP} for the case of GMV system. 

In the present report we intend to construct expansions over the eigenfunctions of the recursion operators related to \eqref{eq:gsys} with arbitrary constant asymptotic values of the potential functions $(u,v)$. In doing this, we are taking into account the whole spectrum of the scattering operator $\tilde{L}^0$, thus extending in a natural way some of the results published in \cite{YanVal2017}. Next, we are showing how these expansions modify depending on the symmetries of the functions we expand. We would like to stress on the following:
\begin{itemize}
\item We shall be dealing with both $\rm GMV_{\pm}$ systems simultaneously;
\item Our approach will be based on the gauge equivalence we mentioned in the above. Consequently, we shall be able to consider general asymptotic conditions --- constant limits $\lim_{x\to \pm\infty}u$ and $\lim_{x\to \pm\infty}v$ ;
\item Our point of view on the recursion operators when reductions are present is somewhat different from that adopted in \cite{GMVSIGMA};
\item We show some new algebraic features in the spectral theory.
\end{itemize}

\section{Gauge-Equivalent Systems}
\label{sec:2}

As mentioned, our approach to the $\rm GMV_{\pm}$ system will be based on the fact that it is gauge equivalent to a GZS system on $\asl(3,\mathbb{C}$) and will follow some of the ideas of \cite{GYaSAM2015,Yan2015SPT}. We  presented our results for the case of the continuous spectrum in \cite{YanVal2017}. Here, we shall include also the discrete spectrum. In fact, we have the following basic result:
\begin{theorem}\label{th:GEq}
The $\rm GMV_{\pm}$ system is gauge equivalent to a canonical {\rm GZS} linear problem on $\asl(3,\mathbb{C})$
\begin{equation}\label{eq:GZSours}
L^0\psi=(\ri\partial_x +q-\lambda J_0)\psi = 0
\end{equation}
subject to a Mikhailov reduction group  generated by the two elements $h_1$ and $h_2$. For a fundamental solution $\psi$ of system \eqref{eq:GZSours} we have:
\begin{equation}\nonumber\label{eq:MGnew}
\begin{array}{l}
{h_1(\psi)(x,\lambda)=\left[Q_{\epsilon}\psi(x,\lambda^{*})^{\dag}Q_{\epsilon}\right]^{-1},
\qquad Q_{\epsilon}=\diag(1,\epsilon,1),\quad \epsilon=\pm 1},\\[4pt]
{h_2(\psi)(x,\lambda)=K\psi(x,-\lambda)K}\, .
\end{array}
\end{equation}
Since $h_1^2=h_2^2=\id$ and $h_1h_2=h_2h_1$ we have again a $\mathbb{Z}_2\times \mathbb{Z}_2$ reduction.
In the above
\begin{eqnarray}\nonumber\label{eq:K}
K=\left(\begin{array}{ccc} 0&0&1\\ 0&1& 0\\1&0&0\end{array}\right)\, .
\end{eqnarray}
\end{theorem}
{\bf Proof:}
Indeed, it is enough to put  $q=\ri\psi_0^{-1}(\psi_0)_x$ where 
\begin{equation}\nonumber\label{eq:psi0}
\psi_0 = \exp\left[- \ri J'\int_{-\infty}^xb(y)\rd y\right] g^{-1}\, .
\end{equation}
In the above expression  $J'=\diag (1,-2,1)$, $b(x)= \ri(\epsilon uu_x^*+vv_x^*)/2$ (note that this expression is real) and $g$, $J_0$ are the same as in \eqref{eq:g-tr}. 
Then $\psi_0$ is a solution to \eqref{eq:GZSours} for $\lambda=0$ and $\rm GMV_{\epsilon}$ is gauge-equivalent to GZS.$\quad\blacksquare$

Also, one gets the following important formulas:
\begin{equation}\label{eq:sympsio}
K\psi_0=\psi_0H, \qquad \psi_0^{-1}K=H\psi_0^{-1}\, .
\end{equation}
In order to continue we shall need some simple algebraic facts.

\section{Algebraic Preliminaries}
\label{sec:3}

The reductions we introduced have clear algebraic meaning: $h(X)=HXH$, $k(X)=KXK$ are obviously involutive automorphisms of the algebra  $\asl(3,\mathbb{C})$ and $\sigma_{\epsilon}X=-Q_{\epsilon}X^{\dag}Q_{\epsilon}$ defines a complex conjugation of the same algebra. As it is known $\asl(3,\mathbb{C})$  is a simple Lie algebra, the canonical choice for its Cartan subalgebra $\mathfrak{h}$ is the subalgebra of the diagonal matrices. It is also equal to $\mathfrak{h}_J=\ker\ad_J$ where $J$ is any diagonal matrix $\diag(\lambda_1, \lambda_2, \lambda_3)$ with distinct $\lambda_i$'s.  In that case we shall call $\mathfrak{h}_J$ \textit{the Cartan subalgebra} and denote it simply by $\mathfrak{h}$ (in particular, we have $\mathfrak{h}_{J_0}=\mathfrak{h}$). More generally, if $S$ is diagonalizable with distinct eigenvalues then $\mathfrak{h}_S=\ker\ad_S$ is also a Cartan subalgebra. We shall denote the projection onto the orthogonal complement $\mathfrak{h}^{\perp}=\mathfrak{h}_J^{\perp}$  of $\mathfrak{h}=\mathfrak{h}_J$ (with respect to the Killing form) by $\pi_0=\pi_J$ when $J$ is diagonal and the projection onto the orthogonal complement $\mathfrak{h}_S^{\perp}$  of $\mathfrak{h}_S$ by $\pi_S$ when $S$ is diagonalizable. One can introduce the system of roots $\Delta$, the systems of positive and negative roots $\Delta_{\pm}$ in a canonical way. The set $\Delta_{+}$ contains $\alpha_1,\alpha_2$ and $\alpha_3=\alpha_1+\alpha_2$, the Cartan-Weil basis shall be denoted by $E_{\pm \alpha_i}, H_1,H_2$ etc. We use the notation and normalizations used in the well known monograph on semisimple Lie algebras \cite{GoGr}. The matrices $H_{1}, H_{2}$ span $\mathfrak{h}$ and the matrices $E_{\alpha}$, $\alpha\in\Delta$ span $\mathfrak{h}^{\perp}$. The complex conjugation $\sigma_{\epsilon}$ defines the real form $\su(3)$ ($\epsilon=+1$)  or  the real form $\su(2,1)$ ($\epsilon=-1$) of  $\asl(3,\mathbb{C})$. If we introduce the spaces
\begin{equation}\label{eq:INsp}
 {\tilde{\g}^{[j]}=\{X: h(X)=(-1)^{j}X\}, \qquad j=0;1~\mod(2)}
\end{equation}
then we shall have the orthogonal splittings
\begin{equation}\label{eq:spltH}
\begin{array}{l}
 {\asl(3,\mathbb{C}) = \tilde{\g}^{[0]}\oplus \tilde{\g}^{[1]}}\, ,\\[4pt]
 {\su(\epsilon) = (\tilde{\g}^{[0]}\cap\su(\epsilon))\oplus ( \tilde{\g}^{[1]}\cap\su(\epsilon))} \; .
\end{array}
\end{equation}
In order to explain our results we shall also need the action ${\cal K}$ of $k:k(X)=KXK$ on the roots: 
\begin{eqnarray}\nonumber
&&{\cal K}(\pm \alpha_{1})={\mp\alpha_2} \; , \qquad {\cal K}(\pm \alpha_{3})=\mp\alpha_3 \; , \qquad {\cal K}(\pm \alpha_{2})=\mp \alpha_1
\end{eqnarray}
so we have {$k(E_{\alpha})=E_{{\cal K}\alpha}$}. We also note that we have the following relations which are used in all calculations:
\begin{equation}\label{eq:f-spaces}
h\circ\ad_S=-\ad_S\circ h \; ,\qquad \sigma_{\epsilon}\circ\ad_S=-\ad_S\circ \sigma_{\epsilon}\, .
\end{equation}
Another issue we must discuss is the relation between $h$ from one side and $\ad_S^{-1}$ and $\pi_S$  from the other.  Here  $\ad_S^{-1}$ is defined only on the space $\mathfrak{h}_S^{\perp}$ but one could extend it as zero on $\mathfrak{h}_S$ which we shall always assume.  One obtains that
\begin{eqnarray}\label{eq:symSh-s}
&&\ad_S^{-1}\circ h=-h\circ \ad_S^{-1},\qquad \pi_S\circ h=h\circ \pi_S\, ,\\
&&\ad_S^{-1}\circ  \sigma_{\epsilon} =- \sigma_{\epsilon}\circ \ad_S^{-1},\qquad \pi_S\circ  \sigma_{\epsilon}=\sigma_{\epsilon}\circ \pi_S\, .
\end{eqnarray}

\section{Recursion Operators of $\rm GMV_{\epsilon}$}
\label{sec:4}
Recursion operators (also called generating operators or $\Lambda$-operators) are theoretical tools that permit:
\begin{itemize}
\item To describe the hierarchies of the nonlinear evolution equations (NLEEs) related to the auxiliary linear problems of GZS type (the AKNS approach \cite{AKNS});
\item To describe the hierarchies of conservation laws for these NLEEs;
\item To describe the hierarchies of compatible Hamiltonian structures of these NLEEs;
\item The expansions over their eigenfunctions permit to interpret the inverse scattering problems for GZS systems as generalized Fourier transforms, {\bf see} \cite{G86,GYa94,IKhKi94};
\item  Recursion operators have important geometric interpretation --- the NLEEs could be viewed as fundamental fields of a Poisson-Nijenhuis structure on the infinite dimensional manifold of "potentials", a concept introduced in \cite{Mag78}.
\end{itemize}
For all these aspects of recursion operators see also the monograph book \cite{GerViYa2008} which contains an extensive bibliography for publications prior to 2008.

The recursion operators $\tilde{\Lambda}_{\pm}$ arise naturally when one tries to find the hierarchy of Lax pairs related to a particular auxiliary GZS linear problem. Assume this problem has the form $\tilde{L}=\ri\partial_x-\lambda S$  where $S$ is in the orbit of the element $J_0$ with no additional assumptions on $S$ and we have that $\tilde L=\psi_0^{-1}L^0\psi_0$ where $L^0$ is a GZS system with $J=J_0$ and $\psi_0$ is a solution to \eqref{eq:GZSours} for $\lambda=0$. We have
\begin{equation}\nonumber
{\tilde{\Lambda}_{\pm}=\Ad(\psi^{-1}_0)\circ \Lambda_{\pm}\circ  \Ad(\psi_0)}
\end{equation}
where $\Lambda_{\pm}$ are the recursion operators for $L^0$, see \cite{Yan93}. The explicit form of $\tilde{\Lambda}_\pm$ is
\begin{eqnarray}\nonumber\label{eq:ROGMV}
&&{\tilde{\Lambda}_\pm({Z})=} \ri {\ad}_S^{-1} {\pi}_S
\left\{ \partial_x{Z} + \frac{S_x}{12}\int\limits_{\pm\infty}^x \langle {Z}, S_y \rangle {\rd} y + \frac{S_{1x}}{4}\int\limits_{\pm\infty}^x
\langle{Z}, S_{1y} \rangle {\rd} y \right\}
\end{eqnarray}
where $S_1=S^2- 2/3$ and $S_{1x}=(S_1)_x$.  $\rm GMV_{\epsilon}$ is a particular case  of a $\asl(3)$ problem so the operators $\tilde{\Lambda}_\pm$ are the recursion operators for $\rm GMV_{\pm}$ system and give the corresponding NLEEs. However, one must be a little more cautious here if one wants to obtain those NLEEs that are compatible with the reduction group. Indeed, the Lax pairs that obey the reductions give hierarchies of equations that have the form:
\begin{eqnarray}\label{eq:NLEEs1}
&&\ad_{S}^{-1}\partial_t S = \sum\limits_{k=0}^r a_{2k}(\tilde{\Lambda}_{\pm})^{2k} \ad_S^{-1}(S_x)+
\sum\limits_{k=1}^m a_{2k-1}(\tilde{\Lambda}_{\pm})^{2k-1} \ad_S^{-1}(S_{1x})
\end{eqnarray}
where $a_i$ are some real constants. We shall not enter in more details here, see \cite{YanVal2017} for this, but one can see that when one considers the hierarchies  of the NLEEs the next equation in the hierarchy is obtained not using $\tilde{\Lambda}_\pm$ but $\tilde{\Lambda}^2_\pm$. So one needs to understand what happens with the expansions that play role of generalized Fourier transform.

\section{Spectral Theory of the Recursion Operators}
\label{sec:5}

The properties of the fundamental analytic solutions (FAS) of the GZS systems play a paramount role in the spectral theory of such systems. In fact, from the canonical FAS (denoted by $\chi^{\pm}$) in canonical gauge one immediately obtains FAS in the pole gauge $\tilde{\chi}^{\pm}$ (with the same analytic properties) \cite{GerViYa2008}. In our case we have $\tilde{\chi}^{\pm}(x,\lambda)=\psi_0^{-1}\chi^{\pm}(x,\lambda)$. The superscripts $\pm$ mean that the corresponding solution is analytic in $\mathbb{C}_{\pm}$ (upper and lower half-plane). For these solutions one has \begin{theorem}\label{fas_sym}
The FAS $\tilde{\chi}^{\pm}(x,\lambda)$ corresponding to the $GMV_{\epsilon}$ system satisfy:
\begin{equation}
Q_{\epsilon}(\tilde{\chi}^{\pm}(x,\lambda^*))^{\dag}Q_{\epsilon}=(\tilde{\chi}^{\mp}(x,\lambda))^{-1},\qquad H\tilde{\chi}^{\pm}(x,\lambda)H = \tilde{\chi}^{\mp}(x,-\lambda)KH\, .
\end{equation}
\end{theorem}

Further,  one builds the so-called adjoint solutions (or generalized exponents) for the GZS systems:
\begin{itemize}
\item GZS system in canonical gauge: ${\bf e}^{\pm}_{\alpha}=\pi_0\chi^{\pm}E_{\alpha}(\chi^{\pm})^{-1}$;
\item GZS system in pole gauge: $\tilde{\bf e}^{\pm}_{\alpha}=\pi_S\tilde{\chi}^{\pm}E_{\alpha}(\tilde{\chi}^{\pm})^{-1}$ .
\end{itemize}
One sees that $\tilde{\bf e}^{\pm}_{\alpha}=\Ad(\psi^{-1}_0){\bf e}^{\pm}_{\alpha}$ and then the fact that they are eigenfunctions of $\tilde{\Lambda}_{\pm}$ and the completeness relations for them become immediate from the classical results for the recursion operators in canonical gauge. Indeed, first
\begin{equation}\nonumber\label{eq:eigentilde}
\begin{array}{c}
{\tilde{\Lambda}_{-}(\tilde{\bf e}^{+}_{\alpha}(x,\lambda))=\lambda \tilde{\bf e}^{+}_{\alpha}(x,\lambda)\, , \qquad \tilde{\Lambda}_{-}(\tilde{\bf e}^{-}_{-\alpha}(x,\lambda))=\lambda \tilde{\bf e}^{-}_{-\alpha}(x,\lambda)}\, ,\\[4pt]\nonumber
{\tilde{\Lambda}_{+}(\tilde{\bf e}^{+}_{-\alpha}(x,\lambda))=\lambda \tilde{\bf e}^{+}_{-\alpha}(x,\lambda)\, , \qquad \tilde{\Lambda}_{+}(\tilde{\bf e}^{-}_{\alpha}(x,\lambda))=\lambda \tilde{\bf e}^{-}_{\alpha}(x,\lambda)}
\end{array}
\end{equation}
and the completeness relations could be written into the following useful form \cite{GYa94}:
\begin{eqnarray}\nonumber\label{eq:expanpoleds}
&&{\delta(x-y)\tilde{P}_0 ={\rm DSC_p}+}\\ \nonumber
&&{\frac{1}{2\pi} \int\limits_{-\infty}^\infty \left[
\sum_{\alpha
\in \Delta_{+}} \tilde{\bf e}^+_\alpha(x,\lambda)\otimes \tilde{\bf e}^+_{-\alpha}(y,\lambda)   -
\tilde{\bf e}^{-}_{-\alpha}(x,\lambda)\otimes\tilde{\bf e}^{-}_{\alpha}(y,\lambda)\right] \rd\lambda}
 \end{eqnarray}
where $\rm DSC_p$ is the discrete spectrum contribution. The second term is the continuous spectrum contribution which we denote by $\rm CSC_p$. Also, in the above 
\begin{eqnarray}\nonumber
&&\tilde{P}_0=\sum_{\alpha\in \Delta}\frac{1}{ \alpha(J_0)}(\tilde{E}_{\alpha}\otimes \tilde{E}_{-\alpha})\, , \qquad \tilde{E}_{\alpha}=\Ad(\psi^{-1}_0)E_{\alpha} = \psi^{-1}_0 E_{\alpha}\psi_0\, ,\\ \nonumber
&& \tilde{\bf e}^{\pm}_{\alpha}=\Ad(\psi^{-1}_0){\bf e}^{\pm}_{\alpha}\,.
\end{eqnarray}
For $\rm DSC_p$, assuming that one has $N^+$ poles $\lambda_i^+$, $1\leq i\leq N^+$ in the upper half-plane  $\mathbb{C}_+$ and $N^-$ poles  $\lambda_i^-$, $1\leq i\leq N^-$ in the lower  half-plane $\mathbb{C}_-$, we get
\begin{eqnarray}\label{eq:DSC}
&&{\rm DSC}_p= \\ \nonumber
&& -\ri \sum\limits_{\alpha\in \Delta_+}\sum\limits_{k=1}^{N^+}\Res(\tilde{Q}^+_{\alpha}(x,y,\lambda) ;\lambda^+_{k})-\ri \sum\limits_{\alpha\in \Delta_+}\sum\limits_{k=1}^{N^-}\Res(\tilde{Q}^-_{-\alpha}(x,y,\lambda) ;\lambda^-_{k})\, ,\\
\nonumber
&&\tilde{Q}^{+}_{\alpha}(x,y,\lambda)=\tilde{\bf e}^{+}_{\alpha}(x,\lambda)\otimes \tilde{\bf e}^{+}_{-\alpha}(y,\lambda),\qquad \ima(\lambda)>0\, ,\\ \nonumber
&&\tilde{Q}^{-}_{-\alpha}(x,y,\lambda)=\tilde{\bf e}^{-}_{-\alpha}(x,\lambda)\otimes \tilde{\bf e}^{-}_{\alpha}(y,\lambda),\qquad \ima(\lambda)<0\, .
\end{eqnarray}

\section{$\Lambda$-Operators and Reductions}
\label{sec:6}
Let us see now the implications of the reductions on the expansions over adjoint solutions. We start with the reduction defined by $h$. Since for the FAS we have the properties stated in Theorem \ref{fas_sym}, for $\beta\in \Delta$ we obtain
\begin{equation}\nonumber\label{eq:symofe-p}
{h (\tilde{\bf e}_{\beta}^{\pm}(x,\lambda))=\tilde{\bf e}^{\mp}_{\mathcal{K}\beta}(x,-\lambda)} \; .
\end{equation}
Changing the variable $\lambda$ to $-\lambda$, taking into account \ that $\mathcal K$ maps the positive roots into the negative ones and vice versa, we obtain after some algebraic transformations
\begin{eqnarray}\nonumber \label{eq:CSCredtenf1}
 &&{{\rm CSC_p}=} {\displaystyle\frac{A_h}{2\pi}\int\limits_{-\infty}^\infty\left[
\sum\limits_{\alpha
\in \Delta_{+}} \tilde{\bf e}^+_\alpha(x,\lambda)\otimes\tilde{\bf e}^+_{-\alpha}(y,\lambda)   -
\tilde{\bf e}^{-}_{-\alpha}(x,\lambda)\otimes\tilde{\bf e}^{-}_{\alpha}(y,\lambda)\right] \rd\lambda}
 \end{eqnarray}
where 
\begin{equation}\label{eq:ah}
{A_h=\frac{1}{2}(\id -h\otimes h)}\, .
\end{equation}
Let us explain what the presence of the ``multiplier'' $A_h$  means. For simplicity let us first assume we have only continuous spectrum. 

Assume $\tilde{Z}(x)$ is such  that $h(\tilde{Z})=\tilde{Z}$ and let us make a contraction first to the right followed by integration over $y$ and to the the left followed by integration over $x$. Then taking into account that $h$ is automorphism and the Killing form is invariant under automorphisms we get  
\begin{eqnarray}\nonumber\label{eq:newexpviaa}
&&{\tilde{Z}(x) = \displaystyle\frac{1}{2\pi} \displaystyle \int\limits_{-\infty}^\infty \left[
\sum\limits_{\alpha
\in \Delta_{+}}\tilde{\bf s}^{\eta}_\alpha(x,\lambda)\mu^{\eta}_{\alpha}  -
\tilde{\bf s}^{-\eta}_{-\alpha}(x,\lambda))\mu^{-\eta}_{\alpha}\right] \rd\lambda}
\end{eqnarray}
where $\eta=+$ ($\eta=-$) depending whether we contract to the left or to the right and  
\begin{eqnarray}\nonumber
&&{\mu^{\eta}_{\alpha}=\langle\langle \tilde{\bf a}^\eta_{-\alpha},[S,\tilde{Z}]\rangle\rangle},\quad {\mu^{-\eta}_{\alpha}=\langle\langle \tilde{\bf a}^\eta_{\alpha},[S,\tilde{Z}]\rangle\rangle}\, ,\\ \nonumber \label{eq:sfun}
&&{\tilde{\bf s}^{\eta}_{\pm \alpha}(x,\lambda)=\displaystyle \frac{1}{2}\left(\tilde{\bf e}^{\eta}_{\pm\alpha}(x,\lambda)+h(\tilde{\bf e}^{\eta}_{\pm\alpha}(x,\lambda))\right)}\, ,\\ \nonumber\label{eq:afun}
&&{\tilde{\bf a}^{\eta}_{\pm \alpha}(x,\lambda)=\displaystyle \frac{1}{2}\left(\tilde{\bf e}^{\eta}_{\pm\alpha}(x,\lambda)-h(\tilde{\bf e}^{\eta}_{\pm\alpha}(x,\lambda))\right)}\, ,
\end{eqnarray}
and for two functions $\tilde{Z}_1(x),\tilde{Z}_2(x)$ with values in $\asl(3)$ we used the notation
$$
\langle\langle\tilde{Z}_1,\tilde{Z}_2\rangle\rangle=\int\limits_{-\infty}^{+\infty}\langle\tilde{Z}_1(x),\tilde{Z}_2(x)\rangle{\rm d} x \; .
$$
If instead of $h(\tilde{Z})=\tilde{Z}$ we assume that $h(\tilde{Z})=-\tilde{Z}$ then in the 
same manner we shall obtain expansions over the functions $\tilde{\bf a}^{\eta}_\alpha$ and the coefficients are calculated via the functions $\tilde{\bf s}^{\eta}_\alpha$.
Since $(\id \pm h)/2$ are in fact projectors onto the $\pm 1$ eigenspaces of $h$
\begin{equation}\nonumber
{h(\tilde{\bf s}^{\eta}_{\pm \alpha}(x,\lambda))=\tilde{\bf s}^{\eta}_{\pm \alpha}(x,\lambda)\; , \qquad h(\tilde{\bf a}^{\eta}_{\pm \alpha}(x,\lambda))=-\tilde{\bf a}^{\eta}_{\pm \alpha}(x,\lambda)}\; .
\end{equation}  
Thus, in case $h(\tilde{Z})=\tilde{Z}$ or $h(\tilde{Z})=-\tilde{Z}$ the expansions could be written in terms of new sets of adjoint solutions, $\tilde{\bf s}^{\eta}_{\pm \alpha}$ or $\tilde{\bf a}^{\eta}_{\pm \alpha}$ that reflect the symmetry of $\tilde{Z}$. For $\alpha\in \Delta_{+}$ one obtains that
\begin{equation}\nonumber\label{eq:asminus}
\begin{array}{c}
\tilde{\Lambda}_{-}(\tilde{\bf s}^{+}_{\alpha}(x,\lambda))=\lambda \tilde{\bf a}^{+}_{\alpha}(x,\lambda)\; , \qquad \tilde{\Lambda}_{-}(\tilde{\bf s}^{-}_{-\alpha}(x,\lambda))=\lambda \tilde{\bf a}^{-}_{-\alpha}(x,\lambda)\; ,\\[4pt]
\tilde{\Lambda}_{-}(\tilde{\bf a}^{+}_{\alpha}(x,\lambda))=\lambda \tilde{\bf s}^{+}_{\alpha}(x,\lambda)\; , \qquad \tilde{\Lambda}_{-}(\tilde{\bf a}^{-}_{-\alpha}(x,\lambda))=\lambda \tilde{\bf s}^{-}_{- \alpha}(x,\lambda)
\end{array}
\end{equation}
and also
\begin{equation}\nonumber\label{eq:asplus}
\begin{array}{c}
\tilde{\Lambda}_{+}(\tilde{\bf s}^{+}_{-\alpha}(x,\lambda))=\lambda \tilde{\bf a}^{+}_{-\alpha}(x,\lambda) \; , \qquad \tilde{\Lambda}_{+}(\tilde{\bf s}^{-}_{\alpha}(x,\lambda))=\lambda \tilde{\bf a}^{-}_{\alpha}(x,\lambda)\; , \\[4pt]
\tilde{\Lambda}_{+}(\tilde{\bf a}^{+}_{-\alpha}(x,\lambda))=\lambda \tilde{\bf s}^{+}_{-\alpha}(x,\lambda) \; , \qquad \tilde{\Lambda}_{+}(\tilde{\bf a}^{-}_{\alpha}(x,\lambda))=\lambda \tilde{\bf s}^{-}_{\alpha}(x,\lambda) \; .
\end{array}
\end{equation}
One sees that the functions in the expansions when we have some symmetry with respect to $h$ are eigenfunctions for  $\tilde{\Lambda}^2_{-}$ ($\tilde{\Lambda}^2_{+}$) with eigenvalue $\lambda^2$.This together with the fact that when recursively finding the coefficients for the Lax pairs one effectively uses $\tilde{\Lambda}^2_{+}$ leads to the interpretation that in case we have $\mathbb{Z}_2$ reduction defined by $h$ the role of the generating operator is played by $\tilde{\Lambda}_{\pm}^2$.

All this happens because of the new form of the expansions, involving the ``multiplier'' $A_h= (\id +h\otimes h)/2$. The point is that the ``multiplier'' $A_h$ has simple algebraic meaning:
\begin{theorem}
The operator $A_h= (\id +h\otimes h)/2$ (acting on $\g\otimes \g$ where $\g=\asl(3,\mathbb{C})$)  is a projector onto the space
\begin{equation}\nonumber 
V=\left(\tilde{\g}^{[0]}\otimes \tilde{\g}^{[1]}\right)\oplus \left(\tilde{\g}^{[1]}\otimes \tilde{\g}^{[0]}\right)\, .
\end{equation}
\end{theorem} Consequently, when for $B\in V$ one makes a contraction (from the right or from the left) with $[S,X]$ where $X$ is in $\tilde{\g}^{[s]}$, then $[S,X]\in \tilde{\g}^{[s+1]}$  and $B\,[S,X]\in \tilde{\g}^{[s]}$.

Let us consider  now the discrete spectrum term more closely. For a GZS system in pole gauge in general position one has $N^+$ poles $\lambda_i^+$, $1\leq i\leq N^+$ in the upper half-plane  $\mathbb{C}_+$ and $N^-$ poles  $\lambda_i^-$, $1\leq i\leq N^-$ in the lower  half-plane $\mathbb{C}_-$. If we have reduction defined by $h$ we see that we must have $N^+=N^-$ since if $\tilde{\bf e}_{\alpha}^{+}(x,\lambda)$ has a pole of some order at $\lambda=\lambda_s^+$ in $\mathbb{C}_+$ then $\tilde{\bf e}^{-}_{\mathcal{K}\alpha}(x,\lambda) $ will have the same type of singularity at $-\lambda_s^+$ in $\mathbb{C}_-$.  In order to simplify the notation we shall put $\lambda_s^+=\lambda_s$, $\lambda_s^-=-\lambda_s$ and $N^+=N^-=N$. Of course, in order to make concrete calculations one needs some assumption on the discrete spectrum. Assume that all the singularities are simple poles, let us consider the contribution from  $\tilde{Q}_{\beta}(\lambda)$ for a fixed $\beta$ and two poles: one pole $\lambda=\lambda_0$ located in $\mathbb{C}_+$  and one pole $\lambda=-\lambda_0$ located in $\mathbb{C}_-$. Then for $\beta\in \Delta$ in some discs around $\lambda_0$ and $-\lambda_0$ we have the Laurent expansions that hold uniformly on $x$:
\begin{eqnarray}
&&\tilde{\bf e}_{\beta}^{+}(x,\lambda)=\frac{\tilde{A}^{+}_{\beta}(x)}{\lambda-\lambda_0}+\tilde{B}^{+}_{\beta}(x)+\tilde{C}^{+}_{\beta}(x)(\lambda-\lambda_0)+\ldots\, ,\\
&&\tilde{\bf e}_{\mathcal{K}\beta}^{-}(x,\lambda)=
\frac{\tilde{A}^{-}_{\mathcal{K}\beta}(x)}{\lambda+\lambda_0}+\tilde{B}^{-}_{\mathcal{K}\beta}(x)+\tilde{C}^{-}_{\mathcal{K}\beta}(x)(\lambda+\lambda_0)+\ldots
\end{eqnarray}
From the properties of the FAS we see that for $\beta\in \Delta$, $h (\tilde{\bf e}_{\beta}^{\pm}(x,\lambda))=\tilde{\bf e}^{\mp}_{\mathcal{K}\beta}(x,-\lambda)$ and consequently
\begin{eqnarray}\label{eq:disspr1-p}
&&h \tilde{A}^+_{\beta}(x)=-\tilde{A}^-_{\mathcal{K}\beta}(x),\quad h \tilde{B}^+_{\beta}(x)=~~\tilde{B}^-_{\mathcal{K}\beta}(x),\quad h \tilde{C}^+_{\beta}(x)=-\tilde{C}^-_{\mathcal{K}\beta}(x)\, .
\end{eqnarray}
For $\alpha\in \Delta_+$ the calculation gives
\begin{equation}
\Res (\tilde{Q}^+_{\beta}(x,y,\lambda);\,\lambda_0)=\tilde{A}^+_{\beta}(x)\otimes \tilde{B}^+_{-\beta}(y)+
\tilde{B}^+_{\beta}(x)\otimes \tilde{A}^+_{-\beta}(y)
\end{equation}
but since the singularities occur in pairs we can combine the contributions from $\lambda_0$ and $-\lambda_0$.  After performing it, we put the formula for $\rm DSC_p$ into a form in which the poles in the upper and lower half-plane play equal role introducing the notation
$$
\lambda_i=\lambda_{i}^+,\qquad \lambda_{i+N}=\lambda_{i}^-=-\lambda_{i}^+,\qquad  1\leq i\leq N\, .
$$
Then one could write
\begin{equation}\label{eq:DSKsym-p}
{\rm DSC_p}=-A_h\sum\limits_{\alpha\in \Delta_+}\sum\limits_{s=1}^{2N}\ri \Res(\tilde{Q}_{\alpha}(x,y,\lambda)\, ;\lambda_{s})\, .
\end{equation}
Now, making contractions to the right (left) by  $[S,\tilde{Z}]$ and integrating one gets the discrete spectrum contribution to the expansion of a given function $\tilde{Z}$:
\begin{eqnarray}\label{eq:dsred}
&&{\rm DSC_p}([S,\tilde{Z}])=\\ \nonumber
&&-2\ri \sum\limits_{\alpha\in \Delta_{+}}  \sum\limits_{k=1}^{N} \tilde{A}_{\epsilon\alpha,k}^{[+;s]}(x)\langle \langle \tilde{B}_{-\epsilon\alpha,k}^{[+;s+1]},[S,\tilde{Z}]\rangle\rangle+\tilde{B}_{\epsilon\alpha,k}^{[+;s]}(x)\langle \langle \tilde{A}_{-\epsilon\alpha,k}^{[+;s+1]},[S,\tilde{Z}]\rangle\rangle
\end{eqnarray}
for $\epsilon=\pm 1$ (depending on what side we contracted). In the above 
\begin{equation}
\tilde{A}_{\beta}^{[+;s]}(y)= \frac{1}{2}(\id+ (-1)^s h) \tilde{A}^+_{\beta}(y)\; , \qquad \tilde{B}_{
\beta}^{[+;s]}(y)=  \frac{1}{2}(\id+ (-1)^s h)\tilde{B}^+_{\beta}(y)
\end{equation}
and $s$ is understood modulo $2$. The action of the recursion operators on the discrete spectrum is not hard to find. For  for $\beta\in \Delta$ and the coefficients of the expansion about $\lambda=\lambda_0$ of  $\tilde{\bf e}_{\beta}^{+}(x,\lambda)$ we get:
\begin{eqnarray}
&& \tilde{\Lambda}_{\pm}\tilde{A}^+_{\beta}(x)=\lambda_0\tilde{A}^+_{\beta}(x) \; , \qquad  \tilde{\Lambda}_{\pm}\tilde{B}^+_{\beta}(x)=\lambda_0\tilde{B}^+_{\beta}(x) +\tilde{A}^+_{\beta}(x)\; .
\end{eqnarray}
Consider now the space $\tilde{V}^+_{\beta}$ spanned by the vectors $\tilde{A}^+_{\beta}, \tilde{B}^+_{\beta}$.  Of course, we  must have $\tilde{A}^+_{\beta}\neq 0$, otherwise there is no singularity. One sees that also $\tilde{B}^+_{\beta}\neq 0$. Next, one checks immediately that the above relations could be true only if $\tilde{A}^+_{\beta}$, $\tilde{B}^+_{\beta}$ are linearly independent, so $\tilde{V}^+_{\beta}$  has dimension $2$ and the matrix of $\tilde{\Lambda}_{\pm}$  in the basis $\tilde{A}^+_{\beta}$, $\tilde{B}^+_{\beta}$ consists of $2\times 2$ Jordan block having $\lambda_0$ on the diagonal.
 
The situation with the spaces  $V^{[+;s]}_{\nu, \beta}$ spanned by the vectors $A^{[+;s]}_{\beta}, B^{[+;s]}_{\beta}\neq 0$ is very similar but slightly more complicated. Indeed,  since $\tilde{\Lambda}_{\pm}\circ h=-h\circ \tilde{\Lambda}_{\pm}$ we obtain that
\begin{eqnarray}
&& \tilde{\Lambda}_{\pm}\tilde{A}^{[+;s]}_{\beta}(x)=\lambda_0\tilde{A}_{\beta}^{[+;s+1]}(x)\, ,\\
&& \tilde{\Lambda}_{\pm}\tilde{B}_{\beta}^{[+;s]}(x)=\lambda_0\tilde{B}_{\beta}^{[+;s+1]}(x) +\tilde{A}_{\beta}^{[+;s+1]}(x)
\end{eqnarray}
and therefore 
\begin{eqnarray}
&& \tilde{\Lambda}_{\pm}^2\tilde{A}^{[+;s]}_{\beta}(x)=\lambda^2_0\tilde{A}_{\beta}^{[+;s]}(x)\, ,\\
&& \tilde{\Lambda}_{\pm}^2\tilde{B}_{\beta}^{[+;s]}(x)=\lambda^2_0\tilde{B}_{\beta}^{[+;s]}(x) +2\lambda_0\tilde{A}_{\beta}^{[+;s]}(x)\, .
\end{eqnarray}
We have  the following options:
\begin{itemize}
\item $\tilde{A}^{[+;s]}_{\beta}\neq 0$. Then one sees that $\tilde{B}^{[+;s]}_{\beta}\neq 0$ and $\tilde{A}^{[+;s]}_{\beta}$, $\tilde{B}^{[+;s]}_{\beta}$ must be linearly independent,  $\tilde{V}^{[+;s]}_{\beta}$  has dimension $2$ and the matrix of $\tilde{\Lambda}_{\pm}^2$  in the basis $2\lambda_0\tilde{A}^{[+;s]}_{\beta}$, $\tilde{B}^{[+;s]}_{\beta}$ consists of $2\times 2$ Jordan block having $\lambda_0^2$ on the diagonal;
\item $\tilde{A}^{[+;s]}_{\beta}=0$. Then if  $\tilde{B}^{[+;s]}_{\beta}\neq 0$ the space  $\tilde{V}^{[+;s]}_{\beta}$  is one dimensional and it is an eigenspace with eigenvalue $\lambda_0^2$$\,$ ;
\item If $\tilde{A}^{[+;s]}_{\beta}=\tilde{B}^{[+;s]}_{\beta}=0$ then $\tilde{V}^{[+;s]}_{\beta}=0$. 
\end{itemize}
In all the cases we see that  for the reduction defined by $h$ the spaces $\tilde{V}^{[+;s]}_{\beta}$ are not invariant under the action of $\tilde{\Lambda}_{\pm}$ but are invariant under the action of $\tilde{\Lambda}^2_{\pm}$. This will happen, of course, when we consider the contribution from all the poles given by the expression \eqref{eq:dsred}.

We have the same effect from the reduction defined by the complex conjugation $\sigma_{\epsilon}$. Both for the continuous and for the discrete spectrum we obtain
\begin{eqnarray}\nonumber
&&{{\rm CSC_p}=A_{\sigma_{\epsilon}}{\rm CSC_p}} \; ,\qquad {{\rm DSC_p}=A_{\sigma_{\epsilon}}{\rm DSC_p}} 
\end{eqnarray}
where
\begin{equation}\label{eq:asigma}
A_{\sigma_{\epsilon}}=\frac{1}{2}(\id-\sigma_{\epsilon}\otimes \sigma_{\epsilon})\, .
\end{equation}
Finally, if the reductions defined by $h$ and $\sigma_{\epsilon}$ act simultaneously then
\begin{eqnarray}
&&{{\rm CSC_p}=A_h A_{\sigma_{\epsilon}}{\rm CSC_p}} \; , \qquad {{\rm DSC_p}=A_h A_{\sigma_{\epsilon}}{\rm DSC_p}}
\end{eqnarray}
where $A_h$ and $A_{\sigma_{\epsilon}}$ are as in \eqref{eq:ah} and \eqref{eq:asigma}.
Note that $A_h$ and  $A_{\sigma_{\epsilon}}$ commute. Of course, in the case of the complex conjugation $\sigma_{\epsilon}$ the role of "symmetric" with respect to the action of $h$ is taken by "real" functions with respect to $\sigma_{\epsilon}$ and the role of "anti-symmetric" with respect to the action of $h$ is taken by "imaginary" functions with respect to $\sigma_{\epsilon}$ and so on. However, both in the case of one $\mathbb{Z}_2$
 reduction ($h$), and in the case of $\mathbb{Z}_2\times \mathbb{Z}_2$ reduction ($h$ and $\sigma_{\epsilon}$) the role played previously by the operators $\tilde{\Lambda}_{\pm}$  is played by $\tilde{\Lambda}^2_{\pm}$ -- the square of these operators.

\section{Conclusion}
\label{sec:7}
We have already discussed that effectively the operators "shifting" the equations along the hierarchies of NLEEs \eqref{eq:NLEEs1} are $\tilde{\Lambda}_{\pm}^{2}$. We have showed that when one uses expansions over adjoint solutions to investigate these evolution equations, then according to the symmetry of the right hand side with respect to $h$ and $\sigma_{\epsilon}$ the expansions modify depending on the symmetries of the functions we expand. So these expansions are over the eigenfunctions of  $\tilde{\Lambda}_{\pm}^{2}$ and in the generalized Fourier expansions the role previously played by  $\tilde{\Lambda}_{\pm}$ is played now by their squares $\tilde{\Lambda}_{\pm}^{2}$.

\section*{Acknowledgement}
The work has been supported by the NRF incentive grant of South Africa and grant DN 02--5 of Bulgarian Fund "Scientific Research".

\end{document}